\documentclass[a4paper]{spie}
\usepackage[]{times}
\usepackage{graphicx}
\title{Reduction of polarimetric data using Mueller calculus applied to Nasmyth instruments\thanks{~~Based on observations obtained at the ESO-NTT at La Silla, Chile (ESO program 80.C-680).}}
\author{Franco Joos\supit{a}, Esther Buenzli\supit{a}, Hans Martin Schmid\supit{a} and Christian Thalmann\supit{a}
\skiplinehalf
\supit{a}Institut f\"ur Astronomie, ETH Z\"urich, 8092 Z\"urich, Switzerland
}
\newcommand{\degr}{^\circ}
\usepackage{txfonts}
\usepackage{amssymb}
\usepackage{lscape}
\usepackage[dvips]{epsfig}
\begin{document}
\maketitle

\begin{abstract}
  
\end{abstract}
We present a method based on Mueller calculus to calibrate linear polarimetric observations. The key advantages of the proposed way of calibration are: (1) that it can be implemented in a data reduction pipeline, (2) that it is possible to do accurate polarimetry also for telescopes/instruments with polarimetric non-friendly architecture (e.g. Nasmyth instruments) and (3) that the proposed strategy is much less time consuming than standard calibration procedures. The telescope/instrument will polarimetrically be described by a train of Mueller matrices. The components of these matrices are dependent on wavelength, incident angle of the incoming light and surface properties. 

The result is, that an observer gets the polarimetrically calibrated data from a reduction pipeline. The data will be corrected for the telescope/instrumental polarisation off-set and with the position angle of polarisation rotated into sky coordinates. Up to now these two calibration steps were mostly performed with the help of dedicated and time consuming night-time calibration measurements of polarisation standard stars.

\keywords{Polarimetry, data reduction, instrumental polarisation, NTT, SOFI}

\section{Introduction}

Light polarisation from astronomical objects (apart from the Sun) has become more and more popular since the first polarimetric observations of comets (Wright\cite{wright81}), of a ``Nebulosity about Nova Persei'' (Perrine\cite{perrine02}), of the moon (Barabascheff \cite{barabascheff27}) or the discoveries of the polarised poles of Jupiter (Lyot\cite{lyot29}). Nowadays, there exist many observatories providing the  possibility for polarimetry, e.g. at the VLT: NACO, FORS1, ISAAC; at the NTT: SOFI and EFOSC2; at Keck 1: LRIS etc. This short list aims not to be complete but rather to show that polarimetry is offered at the world's largest observatories.

However, polarimetric measurements are often affected by telescope and/or instrumental polarisation and a main effort during data reduction is to get rid of these ``disturbing'' effects. 
Especially, observations with Nasmyth instruments are particularly affected by strong and variable telescope polarisation.
The Nasmyth mirror M3 deflects the light by $90\degr$ to the Nasmyth focus and introduces polarisation. Metallic mirrors introduce polarisation dependent on the surface coatings, the incidence angle ($45\degr$ for M3) and the wavelength. For example, an aluminium coated flat mirror with an incidence angle of $45\degr$ introduces up to $5\,\%$ of polarisation at 800\,nm and about $1\,\%$ at 1600\,nm. Although the induced polarisation is intrinsically not dependent on the pointing direction of the telescope (always 45$\degr$ for M3) the measured polarisation directions in a Nasmyth instrument depend strongly on the elevation of the target because the induced polarisation due to M3 is decoupled from the field. Thus, the polarisation breakdown into Stokes components is variable (dependent on pointing direction).

Due to the mentioned reasons but also with the future more complex polarimetric instruments (e.g. polarimetric mode of VLT/SPHERE, see Schmid et al. \cite{schmid06a}) one aims to get the polarimetric data reduction integrated in a pipeline. The pipeline data reduction should be able to correctly handle the telescope/instrument polarisation without the need of time consuming calibration measurements during the night. This can be achieved with a polarimetric instrument model based on Mueller matrices describing all polarimetric effects from the sky to the polarimetric analysing system.
This model needs only few calibration measurements during the night to adjust/monitor the model parameters. 

The next section gives a short description on how to get polarimetric data and on an optimised way of performing polarimetric observations. A good observing strategy is always the first step in achieving good polarimetric data. With the optimal observing technique several instrumental effects can be corrected without applying specific  calibration measurements (self-calibration). The following section describes the standard way of calibrating polarisation measurements. In section \ref{muellersection} the proposed calibration based on a polarisation model described by Mueller matrices is explained in detail and verified with an example measurement. 

The whole paper treats the measurement of linear polarisation only. Circular polarisation is not subject of this publication. ESO instruments/telescopes are used here as examples since the authors have performed several own polarimetric observations at these instruments (e.g. Joos \& Schmid \cite{joos07}, Schmid et al. \cite{schmid06}) Nevertheless, the examples are also valid for similar telescope/instrument configurations.

\section{Different ways to get polarimetric data}

The main optical device to measure polarised light is a linear polariser. This device distinguishes between two orthogonal polarisation directions of light and deflects or absorbs one direction (let's say the horizontal) and transmits the light in the other direction (vertical). The polariser is thus used as the analysing device of polarisation and at the same time it defines the coordinate system of the polarisation directions measured with the instrument. 

A common way to describe polarised light is the Stokes formalism. The light beam is described by a Stokes vector $\vec{S}=(I,Q,U,V)^T$ with the three polarimetric components $Q,U,V$ and the total intensity $I$. The Stokes components are obtained by measuring the intensity of the light after the polariser for different rotational orientations of the polariser. If the default position of the polariser ($0\degr$) transmits light polarised vertically, then the Stokes components are as follows: $Q=I_0 - I_{90}$ and $U=I_{45}-I_{135}$ are the components of linear polarisation, where the indices indicate the rotation angle of the polariser in counter-clock direction when looking towards the sky. $V=I_r - I_l$ is the circular polarisation, where $r$ stands for right and $l$ for left circularly polarised light. The circular polarisation $V$ is only mentioned here for completeness, but will not be followed further. The total intensity is $I=I_0 + I_{90}$ or $ I_{45}+ I_{135}$.

The polarisation degree of the linearly polarised light is defined as:
\begin{equation}
p=\sqrt{p_Q^{2}+ p_U^{2}} \hspace{0.5cm} \textrm{where}\hspace{0.5cm} p_Q = \frac{Q}{I} \hspace{0.5cm}\textrm{and}\hspace{0.5cm} p_U=\frac{U}{I}.
\label{poldeg}
\end{equation}
\noindent 
and the polarisation angle\footnote{The polarisation angle $\theta$ on the sky is usually counted from north over east to south.} as:
\begin{equation}
\theta=\frac{1}{2}\arctan{\left[\frac{p_U}{p_Q}\right]}.
\label{polangle}
\end{equation}

\noindent
There exist two main strategies of getting polarimetric data of an astronomical object, although, the main principle of a differential measurement remains the same:

\begin{itemize}
\item
The first approach is to measure two independent polarisation directions one after the other on the same area of the detector (temporally separated). This is achieved for example by rotating the polarisation analyser by $90\degr$ and thus detecting the two opposite polarisation directions, e.g. firstly $I_0$ and secondly $I_{90}$ for Stokes $Q$ or $I_{45}$ and then $I_{135}$ for Stokes $U$.

\item
The second approach is to get both independent polarisation measurements at once but on different detectors or on  different regions of an imaging detector (spatially separated). For this, one needs a polarising beam-splitter as analyser producing the two beams of orthogonal polarisation at the detector plane (see also Fig. \ref{polprinciple}). The two beams correspond in principle to the two single measurements with a polariser at $0\degr$ and $90\degr$ orientation. Such a system was for example successfully tested and described by Appenzeller \cite{appenzeller67}.
\end{itemize}

\noindent
Both methods have advantages and disadvantages which are described in the next section. There we depict an ``optimised'' way of acquiring polarimetric data while the advantages of the two mentioned methods are combined and the disadvantages are minimised. (A good and extensive tutorial about accurate polarimetry in general and for Nasmyth instruments in particular can also be found in Tinbergen \cite{tinbergen07}).

\section{Optimised way of measuring polarimetric data}\label{optimum}

The advantage of the temporally separated method mentioned above is that the two orthogonal polarisation directions are measured with the same detector element
 and are subject to the same instrument sensitivity. The disadvantage is the time delay during which the atmospheric conditions may change slightly. This may cause a spurious differential signal not due to the target.
On the other hand, the spatially separated method has the advantage of measuring simultaneously both polarisation directions, but in two different channels with different instrument sensitivity. The goal is therefore to combine the advantages of both methods and to eliminate the disadvantages.

This can be reached by the combination of a temporal polarisation exchange and a beam exchange with a polarising beam-splitter (e.g. a Wollaston prism) as polariser, which separates both polarisation directions spatially. After the first exposure the polarisation directions entering the polarising beam-splitter are exchanged by rotating the whole instrument by $90\degr$ or by rotating the polarisation directions of the incoming light by $90\degr$ with a rotatable half-wave plate (HWP) in front of the beam-splitter (this solution is to be favoured!). 

\begin{figure}[h!]
\centering
\epsfig{file=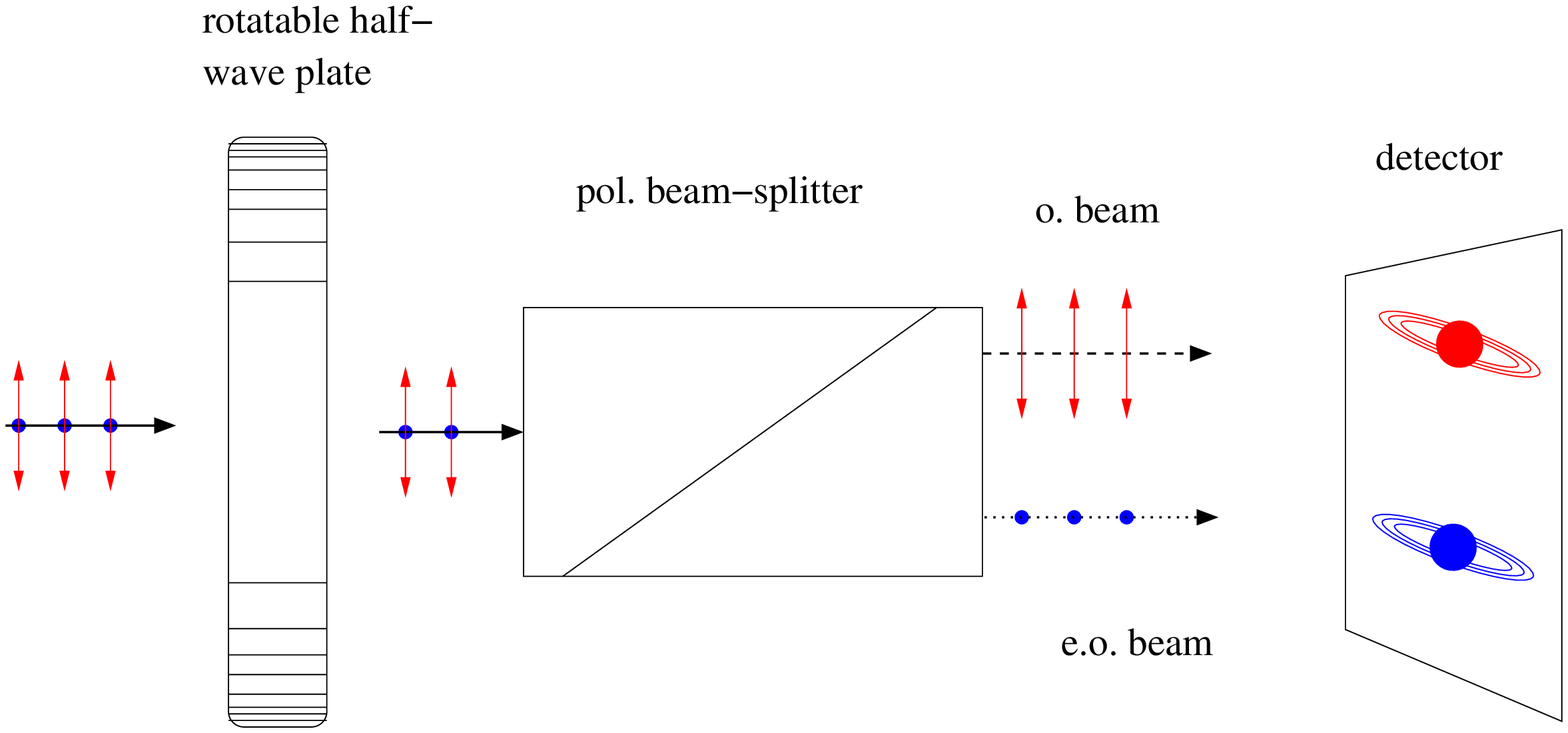,width=10cm} 
\caption{Schematic view of a polarisation analysis system. The light beam enters from the left, passes through the half-wave plate, the polarising beam-splitter and then is split into the ordinary (o.) and extra-ordinary (e.o.) beams and detected at the focal plane. The vertical arrows indicate light polarised vertically and the dots indicate light polarised horizontally.  
}
\label{polprinciple}
\end{figure}

The first exposure is now carried out with the fast optical axis of the HWP at position angle $0\degr$ (e.g. vertically) leading to two spatially separated images $i_{\parallel}(0\degr)$ and $i_{\perp}(0\degr)$
and for the second exposure the HWP is rotated to a position angle of $45\degr$ leading to the next two images $i_{\parallel}(45\degr)$ and $i_{\perp}(45\degr)$. The parallel ($\parallel$) and perpendicular ($\perp$) signs indicate the according polarisation direction split by the polarising beam-splitter, also called ordinary and extra-ordinary beams (cf. Fig. \ref{polprinciple}). The angles in the brackets are the position angles of the HWP. These four images lead to Stokes $Q$. To measure also Stokes $U$ two additional exposures must be performed with HWP angles of $22.5\degr$ and $67.5\degr$ leading to four additional images $i_{\parallel}(22.5\degr), i_{\perp}(22.5\degr), i_{\parallel}(67.5\degr)$ and $i_{\perp}(67.5\degr)$. 

\noindent
In the case where no half-wave plate is available, the whole instrument must be rotated to positions of  $0\degr, 90\degr, 45\degr$ and $135\degr$. This is for example the case at the ESO instrument NTT-SOFI.

The advantage of this measurement technique is that different throughput for the ordinary and extra-ordinary beams of the polarisation analyser will be cancelled if applying the following formula to calculate the normalised Stokes parameters $Q/I$ and $U/I$ (Tinbergen \& Rutten \cite{tinbergen92} ), respectively \footnote{The formula is given for the case of having a half-wave plate. If this is not the case and the whole instrument must be rotated (e.g. SOFI at the ESO-NTT), then the angles in the brackets must be doubled.}:

\begin{equation}
p_Q  = \frac{R-1}{R+1} \hspace{0.5cm} {\rm with} \hspace{0.5cm} R = \sqrt{\frac{I_\parallel (0\degr)/I_\perp (0\degr)}{I_\parallel (45\degr)/I_\perp (45\degr)}}.
\label{tinbergen}
\end{equation}

\noindent
To obtain $p_U$ the $0\degr$ and $45\degr$ must be replaced by $22.5\degr$ and $67.5\degr$, respectively. For spatially resolved data, i.e. extended sources, the reduction formula is applied to the two-dimensional pixel arrays. For these extended sources it is important to achieve a good spatial alignment of the four images. This can be difficult, because some beam-splitters introduce differential chromatic aberrations, so that the images are not exactly congruent (see Schmid et al. \cite{schmid06}). For spectropolarimetry the formula of equation (\ref{tinbergen}) is applied to the spectra (see Joos \& Schmid \cite{joos07}). Important is that the wavelength calibration is accurately done for all images individually.

The described method ``automatically'' calibrates pixel-to-pixel sensitivity variations of the detector and different throughput of the ordinary and extra-ordinary beams of the polarising beam-splitter. This method precisely gives the polarisation at the position of the HWP. However, it does not take into account polarisation effects introduced \textit{in front} of the analysing system, i.e. offsets produced by the telescope itself (inclined mirrors) or by other parts of the instrument (mirrors, lenses, windows, etc.). Further, the polarisation orientation is measured in the coordinate system defined by the polarisation analysis system which in general does not correspond to the coordinate system on the sky. To calibrate these effects, usually calibration measurements of polarimetric standard stars must be applied.

\section{Standard calibration of polarimetric data}

The polarimetric-specific calibration steps include the subtraction of the telescope/instrument polarisation offset and the correction for the zero point of the polarisation angle $\theta$ (relative offset angle between polarisation analyser and the polarisation directions on the sky). The first correction is an additive correction which is applied to the $Q$ and $U$ measurements and the second is a rotational transformation (cross-talk) between $Q$ and $U$. These calibration steps require the measurements of polarimetric standard stars, i.e. zero- and highly polarised standard stars. Those calibration stars can be found in several lists, e.g. in Hsu \& Breger \cite{hsu82}, Whittet et al. \cite{whittet92} or Turnshek et al. \cite{turnshek90} where the polarisation degree $p$ and the polarisation angle $\theta$ (for the highly polarised stars) are given. Since instrumental/telescope polarisation is wavelength dependent it is necessary to measure the standard stars for all used spectral pass-bands.

To determine the polarisation offset of the telescope/instrument a zero polarisation standard star is observed.  After having measured $Q$ and $U$ of the zero polarisation standard star this is compared to the value from the literature (close to zero) and the difference is subtracted from the scientific data.

For determining the zero point of the polarisation angle, highly polarised standard stars must be measured with known polarisation degree $p$ and polarisation direction $\theta$. The measured polarisation direction (according to formula (\ref{polangle})) is then compared to the angle from the literature and a rotational correction is applied to the measured $Q$ and $U$ values.

For simple telescope/instrument configurations like Cassegrain-fed or Nasmyth instruments with only one folding mirror M3 (like in NTT/SOFI) these two types of calibration measurements on polarimetric standard stars are normally sufficient. For the case of at least one additional folding mirror in a Nasmyth instrument, let's say M4, in front of the analyser, the relative position of M3 to M4 affects the telescope/instrument polarisation (like in NACO or ISAAC at the VLT). To calibrate this polarisation offset, which is predominantly dependent on zenith distance, calibration measurements for each zenith distance are required, which is of course very time consuming and thus much less feasible. 

In the following, dedicated polarimetric calibration for two main types of telescope layouts are discussed in more detail: a Cassegrain fed instrument at an equatorial mount telescope and a Nasmyth instrument at an alt-azimuth telescope.
The following two examples are thus only to be understood as typical configurations rather then as recipes for all possible types of telescopes/instruments.

\subsection{Cassegrain at equatorial mounted telescopes}

\paragraph{The telescope polarisation offset} at the highly rotationally symmetric Cassegrain focus is rather easy to determine. One just has to measure one or more zero polarisation standard stars for each requested wavelength, reduce the data in the common polarimetric way, i.e. determine Stokes $Q$ and $U$, and the outcome is already the requested telescope polarisation offset, which is in general very small (~0.1 - 0.5\%) but which can be field dependent. We call the offsets $p_Q^{\rm{off}}$ and $p_U^{\rm{off}}$, respectively. The telescope polarisation offset for this category is not dependent on the pointing direction of the telescope.
Each scientific polarimetric measurement ($p_Q$ and $p_U$) must now be subtracted by the polarisation offset, i.e. $p_Q^{\prime} = p_Q - p_Q^{\rm{off}}$ and $p_U^{\prime} = p_U - p_U^{\rm{off}}$.

\paragraph{The zero point of polarisation direction} is determined by measuring a few highly polarised standard stars with known polarisation degree and position angle. These measured data must of course also first be corrected for the polarisation offset. Then, the polarisation angle $\theta$ is computed according to (\ref{polangle}). The obtained polarisation angle $\theta$ is compared to that from the literature $\theta^{\prime}$ and the scientific data are rotated by the difference $\varphi = \theta^{\prime} - \theta$. The correction for the polarisation zero point is obtained by:
\begin{eqnarray}
p_{Q}^{\prime \prime} &=& p_{Q}^{\prime}\cos{(2\varphi)} - p_{U}^{\prime}\sin{(2\varphi)} \cr
p_{U}^{\prime \prime} &=& p_{Q}^{\prime}\sin{(2\varphi)} + p_{U}^{\prime}\cos{(2\varphi)}. 
\label{qurot}
\end{eqnarray}
Often, it is not clear in which direction $\varphi$ has to be rotated. Therefore, a test with a second or a third highly polarised standard star is recommended. The rotation by $\varphi$ has to be applied to all scientific data and might be wavelength dependent. It is just the relative orientation of the polarisation analysis system versus the sky.

\subsection{Nasmyth instrument at alt-azimuth telescopes }

Instruments attached at the Nasmyth focal plane of alt-azimuth mounted telescopes\footnote{If not mentioned else the examples in this section are for an instrument with only one folding mirror (M3) at the Nasmyth platform A at an alt-azimuth mounted telescope in the southern hemisphere.} are subject to field/pupil rotation dependent on the pointing direction of the telescope, expressed by the parallactic angle $q$ and/or the elevation $e$. The field rotates like $F=q\pm e$ and the pupil with $P=\pm\,e$. The negative signs stand for Nasmyth A platforms\footnote{Definitions and sign conventions can be found in the ESO document ``Field and Pupil Rotation for the VLT Units''\cite{eso}}. 
The field rotation is often stabilised by rotating the instrument attached at a rotator asserting that the sky coordinates are always stable at the detector focal plane.
However, for polarimetric purposes field stabilisation does not suffice. The telescope polarisation mainly caused by the Nasmyth mirror M3 rotates with the pupil, i.e. a field stabilised instrument will measure different $p_Q$ and $p_U$ contributions of the telescope polarisation for different elevations. The offset angle between pupil and field ($P-F$) is, as can be seen from the expressions of field- and pupil rotation, the negative parallactic angle $-q$.

The goal is now to correct the telescope/instrument polarisation offset and polarisation orientation for an arbitrary pointing direction. There are two main approaches to determine the polarisation offset:
\begin{enumerate}
\item For each pointing direction of a scientific target one also measures polarisation standard stars.
\item One measures for one arbitrary pointing direction the zero polarisation standard star(s) and ``normalises'' the measured values to a default telescope/instrument orientation. Later, during data reduction, the normalised telescope polarisation is transformed to the effective scientific pointing directions.
\end{enumerate}

\noindent
The difference lies predominantly in the produced overhead time due to calibration measurements. The first solution is straight forward since one really measures the actual telescope/instrumental polarisation for each pointing direction. There is nothing more to say on this, except that it may consume a lot of observing time and it often would be difficult to find appropriate calibration stars at desired coordinates. The second approach is more elegant but also a coarser approach, since it corrects only the telescope but not the instrumental polarisation. But here one has to notice that if applying the described ``optimised'' way of getting polarimetric data, see section \ref{optimum}, the major part of instrumental polarisation is already compensated by that method. Therefore, the second solution is strongly favoured. In the following, we describe step by step how to execute this method.

One still has to observe zero polarisation and highly polarised standard stars for each spectral domain. The observation of a zero polarisation standard star leads to measured Stokes $p_{Q,1}$ and $p_{U,1}$ according to formula (\ref{tinbergen}). For each observation one needs to know the parallactic angle $q_1$ under which the observation was performed.
Since the measured polarisation is offset by $q_1$ relative to the field (in clock-wise direction) one has to rotate the measured polarisation by $q_1$ in counter-clock direction to get the telescope polarisation locked to the field:

\begin{eqnarray}
p_{Q,0} &=& p_{Q,1}\cos{(-2q_1)} - p_{U,1}\sin{(-2q_1)} \cr
p_{U,0} &=& p_{Q,1}\sin{(-2q_1)} + p_{U,1}\cos{(-2q_1)}. 
\label{qurot2}
\end{eqnarray}

\noindent
To subtract the correct telescope polarisation from a scientific measurement at 
parallactic angle $q$, one has to rotate the telescope polarisation by $q$:
\begin{eqnarray}
p_{Q} &=& p_{Q,0}\cos{(2q)} - p_{U,0}\sin{(2q)} \cr
p_{U} &=& p_{Q,0}\sin{(2q)} + p_{U,0}\cos{(2q)}. 
\label{qurot3}
\end{eqnarray}

\noindent
$p_Q$ and $p_U$ can now be subtracted from the scientific measurements carried out at the parallactic angle $q$. For the case that an exposure is so long that the parallactic angle $q$ is different before and after the exposure one has to use the mean value of $q$. For longer integrations a splitting in shorter integration times would be favourable from a polarimetric point of view but the observer has to disentangle the advantages/disadvantages of splitting also regarding read-out noise, overhead time, etc.

One has to notice that this described calibration method is only a good approximation if the instrument introduces negligible cross-talks and the major part of instrumental polarisation (e.g. different throughput for the ordinary and the extra-ordinary channels) are reduced by the described ``optimised way of measuring polarimetric data''. If this is not the case one has to perform the polarimetric calibration by measuring standard stars for each pointing direction or by polarimetrically modelling the telescope/instrument as proposed in section \ref{muellersection}.

The correction of the polarisation position angle is then again the same procedure like for the previously described case of a Cassegrain fed instrument. Since this correction is only dependent on the relative position of the polarisation analysing system to the sky, which remains constant during a single observing run if field de-rotation is applied, the rotation angle is already obtained after the measurement for only one telescope/instrument orientation.

\section{Polarisation calibration with Mueller calculus}\label{muellersection}

The previously described polarisation calibration is obviously easy for a Cassegrain-fed instrument at an equatorial mounted telescope. However, even for the simplest candidate of a Nasmyth instrument at an alt-azimuth telescope the calibration turns out to be more demanding. An example for such an instrument is SOFI at the Nasmyth focal plane A at the ESO-NTT. 

For even more complex instruments the polarimetric calibration requires much more dedicated calibration measurements. The ESO-VLT second generation instrument SPHERE containing the high precision polarimetric imager ZIMPOL will be such a ``more complex'' instrument with an extreme adaptive optics system, coronagraphy and several folding mirrors (cf. Schmid et al. \cite{schmid06a}). Therefore, there is a strong need for pipeline supported polarimetric calibration and data reduction. 

For this purpose, a polarimetric model of the telescope/instrument is needed. This can best be achieved by the use of Mueller calculus where every single optical element can be polarimetrically described by a $4 \times 4$ Mueller matrix with real elements. This is important for reflecting surfaces like mirrors or for transmitting components like lenses and beam-splitters, but also the rotations applied to the Stokes vector due to e.g. the rotation of a Nasmyth folding mirror can be described by a $4 \times 4$ Mueller matrix. The multiplication of all the single Matrices then leads to a Mueller matrix of the whole telescope/instrument for a given wavelength and a given set-up.

Nevertheless, a model also needs some measurements for improving and fitting the model parameters. Most of the calibration measurements should be obtained during day time, so that only few measurements must still be carried out at night. Hence, the time overhead due to calibration measurements for a complex instrument during the night will  be remarkably diminished when applying the proposed calibration using a telescope/instrument model based on Mueller matrices.

The goal therefore is that the Stokes vector with the two linear parameters $Q$ and $U$ and the intensity $I$ measured with the instrument must only be multiplied by the dedicated Mueller matrix for polarimetric calibration. Therefore, one aims to describe the whole (polarimetric) optical train by Mueller matrices. This must be subdivided into several sub-sections and is shown here step by step at the example of the Nasmyth instrument NTT/SOFI. To do so, we start at the sky with a target star of known polarisation degree $p$ and polarisation direction $\theta$. 

Note that the following matrices are not accounting for the common intensity loss described by the reflectivity factor.

\subsection{Transformation of polarimetric sky coordinates into ``telescope coordinates''}

A point-like object in the sky is polarimetrically described by its polarisation degree $p$ and the polarisation angle $\theta$ counted from north over east to south with $\theta \in  [0,180\degr)$. These parameters must be transformed into a Stokes vector $\vec{S}_{0} = (I_0,Q_0,U_0,V_0)^T$ on the sky to be handled by Mueller matrices. This can be achieved by using the equations for the polarisation degree (\ref{poldeg}) and the polarisation angle (\ref{polangle}).
The $Q$ direction shall be defined in north to south orientation, whereas the $U$ direction is in north-east to south-west direction. 

It is important to orientate the sky Stokes vector $\vec{S}_0$ in the appropriate orientation relative to the Nasmyth folding mirror M3, i.e. the positive $Q$ direction must be perpendicular to the following plane of scattering, which is defined by the incident and the reflected beam of light at M3. From a polarimetric point of view M1 and M2 have only a minor importance due to rotational symmetry and we neglect their effect particularly since it is an even number of mirrors neutralising the sign flip in $U$ (and $V$).

The decisive rotation angle of the sky Stokes vector relative to M3 is described by the parallactic angle $q$, which is dependent on the telescope latitude $\phi$, the hour angle $h$ and the zenith distance $z$:

\begin{equation}
q =\arcsin \left[ \frac{\cos \phi \cdot \sin h}{\sin z}\right].
\label{parallangle1}
\end{equation}

\noindent
The rotation is described by a rotation-Mueller matrix $\mathbf{R}(q)$. Applying this matrix to a Stokes vector will rotate the vector by an angle $q$ in clockwise-direction:

\begin{equation}
{\bf R}(q)=
\left(
\begin{array}{rrrr}
1&0&0&0\\
0&\cos (2q) &\sin (2q) &0\\
0&-\sin (2q) &\cos (2q) &0\\
0&0&0&1
\end{array}
\right).
\label{rotmatrix}
\end{equation}

\noindent
After the rotation by the parallactic angle $q$ the original Stokes vector $\vec{S}_0$ is transformed into $\vec{S}_1 = \mathbf {R}(q)\vec{S}_0$.

\subsection{Polarimetric impact of the Nasmyth folding mirror}

Each oblique reflection of light on a metallic surface, like a mirror, changes its polarimetric properties, i.e. instrumental polarisation and cross-talk between different Stokes parameters is introduced. The reflection on a mirror can polarimetrically be described by a dedicated Mueller matrix. Again the matrix is defined depending on the coordinate system, i.e. on the directions of $Q$ and $U$.

For this purpose the Mueller matrix of the optical element must be defined in a dedicated coordinate system, normally in a way that the positive $Q$ direction is perpendicular to the plane of scattering.

The Mueller matrix of a metallic mirror has the following shape (Collet \cite{collet93})\footnote{The original $M_{33}$ element in Collet\cite{collet93} has the opposite sign!}:

\begin{equation}
\label{muellermatrixmirror}
{\bf M}(\vartheta,\lambda)=\frac{1}{2}
\left(
\begin{array}{rrrr}
1+\rho^{2} & 1-\rho^{2} & 0 & 0\\
1-\rho^{2} & 1+\rho^{2} & 0 & 0\\
0 & 0 & -2\rho\,\cos\delta & -2\rho\,\sin\delta\\
0 & 0 & 2\rho\,\sin\delta & -2\rho\,\cos\delta
\end{array}
\right),
\end{equation}
where $\rho$ and $\delta$ are dependent on the surface material expressed by the complex refraction index $n_{c}=n-i * k$ and the incident angle $\vartheta$ onto the mirror (which is always $45\degr$ for the Nasmyth folding mirror). The expressions $\rho$ and $\delta$ are given by the following terms:

 \begin{eqnarray*}
 \label{muellermatrixmirrorparameters1}
  \rho ^{2} &=& \frac{\sqrt{p^{2} + q^{2}} + s^{2} - 2sr_{+}}{\sqrt{p^{2} + q^{2}} + s^{2} + 2sr_{+}}, \\
\tan\delta &=& \frac{2sr_{-}}{\sqrt{p^{2} + q^{2}} - s^{2}},
 \end{eqnarray*}

\noindent 
where

\begin{eqnarray*}
\label{muellermatrixmirrorparameters2}
p &=& n^{2} - k^{2} - \sin^{2}\vartheta, \\
q &=& 2nk, \\
r_{\pm} &=& \frac{1}{\sqrt{2}}\left(\pm p + \sqrt{p^{2} + q^{2}}\right)^{1/2}, \\
s &=& \sin\vartheta\, \tan\vartheta\,.
\end{eqnarray*}

\noindent
The wavelength dependent refraction indices must be taken from optical tables, e.g. Winsemius et al.\cite{winsemius76}. Note, that mostly only indices for pure metal (e.g. aluminium) are available and one has to feed the model with these approximative parameters although real mirrors will be covered by coatings, oxide layers or dirt changing the polarimetric behaviour. The goal is later to use the calibration measurements to adjust the refraction indices.

The multiplication of the mirror matrix (\ref{muellermatrixmirror}) with $\vec{S}_1$ leads to the Stokes vector in the coordinate system of mirror M3, called $\vec{S}_2$. This is rotated into the system of the Nasmyth platform by an additional rotation $\mathbf{R}(a)$ where $a$ is the altitude angle. $\vec{S}_3=\mathbf{R}(a) \, \vec{S}_2$. Whether the rotation is in clock- or counter-clock direction (positive $a$ or negative $a$) depends on whether the instrument is on platform A or B. For SOFI, which is on platform A, the rotation is in clockwise direction, thus positive $a$.

\subsection{Description of the Nasmyth instrument}

Now, the Stokes vector $\vec{S}_3$ is in the coordinates of the Nasmyth platform. But, since the instrument itself is in addition rotated by an angle $\beta$ (field de-rotation) and the analysing system is fixed to the instrument, the analysis of $\vec{S}_3$ is affected by this rotation. Thus, $\vec{S}_3$ is further rotated by an angle $\beta$: $\vec{S}_4=\mathbf{R}(\beta)\, \vec{S}_3$. Again, whether $\beta$ is counted positive or negative depends on the rotation direction of the instrument. In addition, an offset angle can be added to $\beta$, since the position of the Wollaston inside the instrument can be rotated by a fixed angle relative to the instrument.

Finally, also the instrument itself can be described by a dedicated wavelength dependent Mueller matrix $\mathbf{M}_{\rm i}(\lambda)$. With this instrument Mueller matrix additional optical components like entrance windows, lenses, filters etc. or especially folding mirrors (like in VLT/NACO) would be described. For the case of NTT/SOFI, which is a rather straight forward Nasmyth instrument, its Mueller matrix can be approximated by the $4\times 4$ identity:

\begin{equation}
\label{muellermatrixinstrument}
{\bf M}_{\rm i}=
\left(
\begin{array}{rrrr}
1 & 0 & 0 & 0 \\
0 & 1 & 0 & 0 \\
0 & 0 & 1 & 0\\
0 & 0 & 0 & 1 
\end{array}
\right).
\end{equation}

\noindent
Thus, from the original Stokes vector on the sky $\vec{S}_0$ a modified vector $\vec{S}$ is measured with SOFI:

\begin{equation}
\vec{S} = \underbrace{\mathbf{M}_{\rm i}\, \mathbf{R}(\beta)\,\mathbf{R}(a)\,\mathbf{M}(\vartheta,\lambda)\,\mathbf{R}(q)}_{\mathbf{T}_{\rm ti}}\,\vec{S}_0 = \mathbf{T}_{\rm ti}\,\vec{S}_0,
\label{telescopetrain}
\end{equation}
where $\mathbf{T}_{\rm ti}$ describes the Mueller matrix of the whole telescope and instrument.

\subsection{Calibration of the measurement}
Now, one is interested in the polarimetric signal of the target on the sky. If the Mueller matrix of the telescope $\mathbf{T}_{\rm ti}$ is known, it can be inverted and multiplied with the measured Stokes vector $\vec{S}$ to obtain the original Stokes vector on the sky $\vec{S}_0$ free from telescope polarisation, cross-talk effects due to mirror M3 and rotated into the appropriate coordinate system of the sky:
\begin{equation}
\vec{S}_0 = \mathbf{T}^{-1}_{\rm ti}\,\vec{S}.
\label{inversetelescope}
\end{equation}

\begin{figure}[h!]
\centering
\epsfig{file=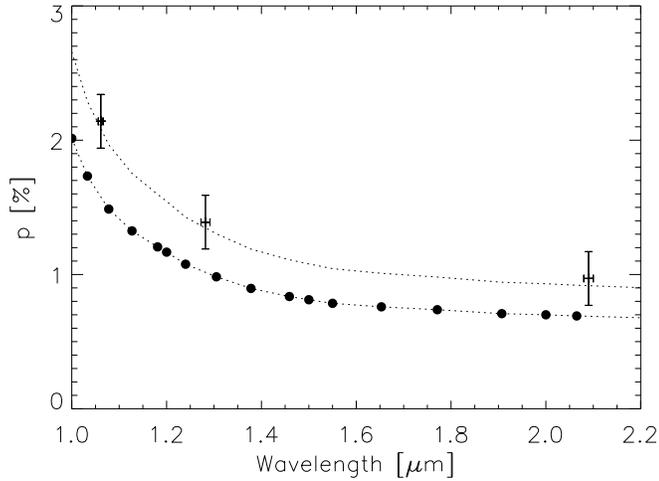,width=10cm} 
\caption{Wavelength dependent instrumental polarisation of the aluminium coated Nasmyth mirror (M3). The filled dots are the calculated polarisation values for pure aluminium (refractive indices from lab) and the lower curve is a best fit to it. The upper curve is the same as the lower but with adapted refraction indices accounting for oxide layers to fit the three measurements (indicated with the error bars).}
\label{obsvsmodel}
\end{figure}

\noindent
With our linear polarimetric measurements performed at the near-infrared instrument SOFI at the ESO 3.5m NTT we tested our model by applying the proposed Mueller calculus to our standard star observations. In Fig. \ref{obsvsmodel}, we show our calculated telescope polarisation caused by M3 for a number of wavelengths (filled dots). For these calculations we used the refraction indices of bulk aluminium. Then we compared the calculated to the measured data points and realised that the calculated polarisation was too low for the whole spectral range. We fitted the complex refraction index $n_c = n - i*k$ so that the calculated polarisation matches the measured data. For this purpose, the real ($n$) and the imaginary ($k$) parts of the refraction indices are lowered by 25\%, which lies within the realms of possibility, according to Smith et al. \cite{smith85}. This deviation of refraction indices is due to the fact that the ``theoretical'' values firstly implemented in the model are gained from measurements under vacuum and laboratory conditions, whereas the measurements of the polarimetric standard stars are performed with the mirrors exposed to air. Due to the oxygen in the air the bulk aluminium immediately forms an aluminium-oxide layer (Al$_2$O$_3$). This layer protects the underlying aluminium from further oxidation but on the other hand has an impact on the polarisation behaviour of the mirror compared to bulk aluminium.

\subsection{A fully calculated example}
As example of reliability we demonstrate our model by means of a highly polarised standard star (HD150193). The star was observed at parallactic angle $q=111.2\degr$ and elevation $e=31.3\degr$. The literature values for the polarisation degree and polarisation angle are $p=3.27\%$ and $\theta=57\degr$ in the J band (Whittet et al.\cite{whittet92}). This corresponds (by applying equations (\ref{poldeg}) and (\ref{polangle})) to a normalised linear Stokes vector
 
\begin{equation}
\vec{S}_0 =
\left(
\begin{array}{r}
1 \\
-0.0133 \\
0.0298 \\
0  
\end{array}
\right).
\end{equation}

\noindent 
The directly measured values (without correcting for polarisation offset and polarisation angle) lead to 

\begin{equation}
\vec{S} =
\left(
\begin{array}{r}
1 \\
-0.0262 \\
0.0207 \\
0  
\end{array}
\right),
\end{equation}

\noindent
for a narrow band filter centred at 1.19\,$\mu$m. 

Our computed Mueller matrix for the telescope/instrument at $q=111.2\degr$, $e=31.3\degr$ and at 1.2\,$\mu$m with adapted refraction indices for Aluminium has the following appearance:

\begin{equation}
\label{muellermatrix}
{\bf T}_{\rm ti}(q,e,\lambda)=
\left(
\begin{array}{rrrr}
0.9848  &  -0.0112  &  -0.0103  &  0.0000\\
-0.0112  &   0.9794  &  0.0059 &     0.1024\\
-0.0103  &  0.0059  &    0.9783  &   -0.1121\\
0.0000   &   0.1024  &   -0.1121  &   -0.9729
\end{array}
\right).
\end{equation}

\noindent
Now, one only has to invert the matrix $\mathbf{T}_{\rm ti}$ from above and multiply the inverted matrix $\mathbf{T}^{-1}_{\rm ti}$ to the measured Stokes vector $\vec{S}$ leading to the fully calibrated Stokes vector $\vec {S}_{0^{\star}}$ of the highly polarised standard star HD150193:

\begin{equation}
\label{muellermatrixinvers}
{\bf T}^{-1}(q,e,\lambda) =
\left(
\begin{array}{rrrr}
1.0157  &   0.0116  &    0.0106 &  0.0000 \\
     0.0116 &       1.0101  &   0.0061  &     0.1056 \\
     0.0106  &   0.0061  &      1.0090  &    -0.1156\\ 
   0.0000  &     0.1056  &    -0.1156  &     -1.0034
\end{array}
\right)
\end{equation}

\noindent
and

\begin{equation}
\vec{S_{0^{\star}}} = \mathbf{T}^{-1} ~ \vec{S} = \left(
\begin{array}{r}
1.0000  \\
-0.0145 \\
0.0308 \\
-0.0051  
\end{array}
\right).
\end{equation}

\noindent
Comparing $\vec{S}_{0^{\star}}$ to the values from the literature $\vec{S}_0$ shows that the difference is well within the tolerances of the measurements which are between $0.1\%$ and about $0.2\%$.

\begin{equation}
\vec{S}_0 - \vec{S}_{0^{\star}} =
\left(
\begin{array}{r}
0 \\
0.0012 \\
-0.0010 \\
0.0051  
\end{array}
\right).
\end{equation}

\noindent
The Stokes-$V$ component is also given here, but this has no relevance, since Stokes-
$V$ was not measured.

\section{Conclusions}

Polarimetry is a powerful technique taking into account also the orientation of the electric field and not only the intensity of light. Nevertheless, polarimetry is rather a niche science. A reason for this is among others the fact that polarimetric measurements are often ``contaminated'' by telescope/instrumental effects which are demanding to calibrate. 

To get good polarimetric observations we strongly recommend the described technique with a combination of temporal and spatial differential measurements (see section \ref{optimum}). To benefit at most from the advantages of this technique a rotatable half-wave plate in front of the polarisation analysis is very important (not offered in NTT/SOFI). If it is not provided, the whole instrument must be rotated to measure at different polarisation angles. This has strong impacts on the polarimetric precision, especially for extended objects, since different detector zones are illuminated differently due to rotating. Many rotations of the whole instrument by large angles in short time may also have an impact onto sensitive parts of the instrument itself, like e.g. the cryogenic system. In addition dithering becomes difficult since the orientation relatively to the occulting masks has to be taken into account. 
Nevertheless, in several instrument hand books (like for the NTT/SOFI) a different and less satisfactory observing technique is proposed which leads to less precise polarimetric measurements.

With our method of calibrating polarimetric measurements by means of Mueller calculus we have demonstrated that polarimetric calibration and data reduction can be implemented into a pipeline and hence, make polarimetric observations more attractive. We have implemented a very simple model for describing the polarimetric properties at NTT/SOFI. This model could also be upgraded by taking into account all optical devices instead only the folding mirror M3. On the other hand we reached with the simple model already an accuracy which is well inside the error of the measurements.

We would therefore strongly encourage to implement the polarimetric data reduction and calibration in a pipeline and benefit not only from the pipeline service itself but also reducing much overhead time up to now ``wasted'' for calibration measurements.

\end{document}